\documentclass[preprint2]{aastex}
\usepackage{graphics}
\def\Lsun{$L_\odot$} 
\def\Msun{$M_\odot$}
\def\msun{$M_\odot$}

\def\teff{$T_{\rm eff}$} 
\def\0BMV{$(B-V)_{\rm 0}$}

\def\feh{{\rm [Fe/H]}}

\def\ocen{$\omega$~Cen}
\def\simgt{\lower.5ex\hbox{$\; \buildrel > \over \sim \;$}} 
\def\simlt{\lower.5ex\hbox{$\; \buildrel < \over \sim \;$}}
\begin{document} 
%
\title{A helium spread among the main sequence stars in  NGC 2808}
\author{F. D'Antona\altaffilmark{1}, M. Bellazzini\altaffilmark{2}, 
V. Caloi\altaffilmark{3}, F. Fusi Pecci\altaffilmark{2}, S. Galleti\altaffilmark{2}, 
R.T. Rood\altaffilmark{4}} 
%
%
\affil{\altaffilmark{1}INAF - Osservatorio Astronomico  di Roma, via Frascati
33, 00127 Roma, Italy; dantona@mporzio.astro.it}
\affil{\altaffilmark{2}INAF - Osservatorio Astronomico  di Bologna, via
Ranzani 1, 40127 Bologna, Italy, michele.bellazzini@bo.astro.it,  
flavio.fusi.pecci@bo.astro.it}
\affil{\altaffilmark{3}
INAF - Istituto di Astrofisica Spaziale e Fisica Cosmica, Via 
Fosso del Cavaliere, I-00133 Roma, Italy; vittoria.caloi@rm.iasf.cnr.it }
\affil{\altaffilmark{4}
Department of Astronomy, University of Virginia, 
Charlottesville,VA 2903-0818, USA
rtr@virginia.edu}
 
\begin{abstract}

We have made a detailed study of the color distribution of the main sequence of 
the Globular Cluster NGC 2808, based on new deep  HST-WFPC2 photometry of a 
field in the uncrowded outskirts of the cluster.  The observed color 
distribution of main sequence stars is not gaussian and  is wider than expected 
for  a single stellar population, given our (carefully determined) measurement 
errors.  About 20\% of the sample stars are much bluer than expected and are 
most plausibly explained as a population having a much larger helium abundance
than the bulk of the main sequence. 
Using synthetic CM diagrams based on new stellar models we estimate that the 
helium mass fraction of these stars is $Y\sim 0.4$.  The newly found anomaly on 
the main sequence gives credit to the idea that GCs like NGC 2808 have suffered 
self--enrichment,  and that different stellar populations were born from the 
ejecta of the  intermediate mass asymptotic giant branch (AGB) stars of the 
first generation. 

Enhancement and spread of helium among the stars in NGC~2808 have been recently 
suggested as a simple way to explain the very peculiar morphology of its 
horizontal branch. We find that if in addition to the $Y=0.40$ stars, roughly 
30\% of the stars have $Y$ distributed between 0.26--0.29 while 50\% have 
primordial $Y$, this leads to a horizontal branch morphology similar to that 
observed. In this framework, three main stages of star  formation are 
identified, the first with primordial helium content $Y\simeq  0.24$, the 
second one born from the winds of the most massive AGBs of the first stellar 
generation ($\sim 6-7$\msun),  having $Y\sim 0.4$, and a third one born from 
the matter ejected from less massive AGBs ($\sim 3.5-4.5$\msun)  with $Y\sim 
0.26-0.29$. There could have been a long hiatus (several 10$^7$yr), between the 
second and third generation, in which no star formed in the protocluster. We 
suggest that, during this period, star formation has been inhibited 
by the explosion of late Supernovae II deriving from binary evolution.

\end{abstract}

\keywords{globular clusters: general --- globular clusters: formation 
--- globular clusters: individual NGC 2808 --- Stars: main sequence}

\section{Introduction}

Chemical inhomogeneities in globular cluster (GC) stars are giving new evidence 
on the formation and evolution of these most ancient stellar systems.  Self--
enrichment mechanisms are being examined as some of the possible---and most 
plausible---causes of the abundance spreads seen in many GCs. In these 
clusters, light elements that are susceptible to abundance changes from proton-
capture reactions, such as the pp, CN, ON, NeNa, and MgAl cycles, exhibit star-
to-star abundance variations, far in excess of the modest variations seen in 
halo field stars---see, e.g., \citet{sne99,sne00,grat-araa}. In some cases 
these abundance spreads are seen in stars near the turnoff and along the 
subgiant branch \citep[e.g.,][]{gratton2001}, suggesting that these anomalies 
arise from some process of self-enrichment occurring at the first stages of 
cluster life. This is generally attributed to pollution from the winds of the 
massive AGBs \citep[e.g.][]{cottrell-dacosta, dgc1983,dantona2003}, whose 
convective envelopes are subject to hot bottom burning (HBB) 
\citep[e.g.][]{ventura2001, ventura2002}. So these envelopes are an ideal 
environment to manufacture elements through nuclear reactions in which proton 
captures are involved.

While problems persist in obtaining a quantitative reproduction of the observed 
abundance spreads \citep{denis2003, ventura2004, fenner2004, ventura-dantona2005},
a high helium abundance in the ejected matter is a robust 
prediction of the models \citep{ventura2001,ventura2002}. \cite{dantona2002} 
have shown that the hypothesis that GC stars were formed in two different star 
formation events can explain the presence of long blue tails in the horizontal 
branches (HB) in GCs with large abundance anomalies \citep{cat1995}. In their 
scenario the bluer HB stars are born from the helium rich stellar population. 
In addition, \cite{dantona-caloi2004} have shown that the peculiar bimodal 
distribution of stars in the cluster NGC 2808 can be explained in the same way, 
by assuming that the cluster contains today about half stars belonging to the 
first stellar generation, and about a half born from the AGB winds, whose 
initial helium abundance is larger than $Y \simeq 0.27$ and extends up to 
$Y\simeq0.35$.

The suggestion that helium is a possible second parameter for globular cluster 
stars \citep[see][for a general overview]{rood73, rc89, fpb97} has been 
reinforced by the discovery that the complex object \ocen\ has at least two 
different components of the main sequence (MS) and multiple turnoffs, 
\citep{bedin2004, ferraro2004} and that the bluer MS can be explained by a 
metal enriched and helium enriched population \citep{norris}. 

In contrast to \ocen, NGC~2808 does not show a metallicity spread 
\citep{cbc04}. The comparison of isochrones differing in helium for the turnoff 
(TO) and MS in \cite{dantona-caloi2004} showed that it is not easy to dismiss 
the hypothesis of differences in helium content among NGC~2808 stars. Indeed 
they predicted that one could observe a color bifurcation in the MS and/or in 
the subgiant branch, if there are at least two homogeneous groups of stars with 
different helium abundance. To investigate this possibility we have used 
HST/WFPC2 photometry of a field in the outskirts of NGC~2808 to obtain a deep, 
high precison Color-Magnitude diagram of the main sequence. Since we were 
searching for a spread in the sequences we also performed an extensive error 
analysis. 

We find that $\sim 20$\% of the MS stars of the cluster lie to the blue of the 
bulk of the MS. The blue outliers must be due to an intrinsic difference 
between the stars (Sect.~2). In Sect.~3 and 4 we suggest an interpretation of 
these anomalous MS stars {\em and} of the HB morphology of the cluster within 
the context outlined above.  In Sect.~5 we discuss and summarize the results.

\section{Observational results}

We reduced a dataset of four $t_{\rm exp}=900$ s plus three $t_{\rm
exp}=100$ s F555W images and four $t_{\rm exp}=700$ s plus three
$t_{\rm exp}=120$ s F814W images, obtained with the WFPC2 on board of
HST.  The images were acquired adopting a dither pattern that allows a
virtually total cleaning of cosmic ray hits and chip defects. The
field is located at $\simeq 4.5\arcmin$ from the center of the
NGC~2808 (GO06804). The field samples the outskirts of the cluster, at
$\sim 17$ core radii ($r_c$) and $\sim 6$ half-light radii from the
center \citep{trag} and it appears relatively uncrowded when imaged
with the superb spatial resolution of the WFPC2. This allowed us to obtain
excellent photometry down to the faintest detectable stars, since
sources confusion were not a concern.  The pre-reduced (flat,~bias)
images provided by the STScI were reduced individually with HSTPHOT
\citep{dol}, a Point-Spread- Function fitting package specifically
suited for WFPC2 images. HSTPHOT automatically provides accurately
calibrated and CTE-corrected photometry. In the following we will only
refer to the photometry calibrated in the Johnson-Kron-Cousin system,
hence in $V$ and $I$ instead of F555W and F814W.\footnote{The results we
present also hold if we maintain the flight system colors, that is,
they are not an artifact of the color transformations.} The adopted
threshold for source detection on the frame was 3 $\sigma$ above the
background. For each filter, the four catalogs from the
long-exposure images were cross-correlated and the average of the
individual measures was taken as the final magnitude estimate, and the
standard deviation was taken as an estimate of the photometric
error. Only stars that were detected and successfully measured in at
least 3 images were retained in the final catalog of average
magnitudes.  We proceeded in the same way for the short-exposure
images, in this case the retained sources were present in at
least two images.  The $V$ and $I$ catalogs of the long and short
exposures were then coupled and finally the $V,~I$ long and short
catalogs were merged after a careful relative photometric
calibration. In the final global catalog all the sources with
$V>20.0$ are from the long exposures, those with $V\le 20.0$ are from
the short ones. Note that the results presented in the following
utilize only stars with $V>20.0$, hence extracted from the homogeneous
set of the long-exposure images.  Note also that the selection
criteria described above ensure that the final dataset is essentially
free from spurious sources, since only stars with at least three
measures in $V$ and 3 measures in $I$ are included in the final catalog,
for $V>20.0$.

\subsection{Color Magnitude Diagram and artificial stars experiments}

The Color Magnitude Diagrams (CMD) obtained from each WFPC2 camera are
shown separately in Fig.~1. The points of the overplotted ridge-lines
have been obtained as 2$\sigma$ clipped averages in color over $0.2$
magnitude boxes, as described in \citet{f99}. For each of the four
subsamples (PC, WF1, WF2, WF3) we derived a separate ridge-line to
account for any subtle camera-to-camera systematic difference in the
photometry. Such effects typically amount to $\pm 0.01$ mag.  A ridge
line for the total sample has also been derived with the same
tecnique.

The CMDs are clearly dominated by the narrow and well defined cluster
sequences: the Sub Giant Branch (SGB) and the base of the Red Giant
Branch (RGB) for $V<19.5$, the Turn Off (TO) point around $V\simeq
20.0$ and the Main Sequence going from $(V,~V-I) \simeq (20.0, 0.8)$
to $\simeq(26.0, 2.0)$. A handful of foreground Galactic stars is
visible in the CMDs of the WF cameras, a couple of magnitudes above
the cluster MS. A sparse sequence running parallel to the MS, $\sim
0.1-0.2$ mag to the red of it, may be due to a mix of blended MS stars
and real binary systems \cite[see][and references therein]{n288}.

On the other hand, there is a sizable fraction of MS stars scattered
to the blue of the ridge lines.  No obvious observational effect can
possibly push a cluster star in this direction. It is worth noting
that while HSTPHOT should be the ideal choice for the present
application, we repeated the data reduction using DoPhot
\citep{dophot}, following the prescriptions by \citet{n288}, and
Sextractor \citep{sex}: the unexpected blue stars appeared the same in
the CMDs obtained with these codes, hence their anomalous color cannot
be due to some subtle effect associated with the reduction process.
These blue MS stars are the main subject of the following subsections.

To accurately characterize all the effects due to the observations
plus data reduction processes and to have an idea of the completeness
of our sample we performed an extensive set of artificial star
experiments on one $V$, and one $I$ images from the long exposure set.
The artificial stars were distributed according to the observed
spatial distribution and luminosity function, as described in
\citet{dol}, and were added simultaneously in the $V$ and $I$ frame at
the same position, hence artificial stars have a properly defined
$V-I$ color \cite[as in][]{n288}. In several runs we accumulated a
total of 91,680 artificial stars, 30,221 of which have input
magnitudes and colors within $\pm 0.15$ mag of the cluster ridge-line
(hereafter Similar Sample).  We used the latter subsample to study in
more detail the MS stars we are interested in.  The average of $V_{\rm
output}-V_{\rm input}$ and of $(V-I)_{\rm output}-(V-I)_{\rm input}$
computed in $\sim 0.5$ mag boxes were taken as the typical photometric
errors in magnitude and color, respectively, and are plotted as error
bars in the various panels of Fig.~1. The average errors derived from
the artificial star experiments are in excellent agreement with the
standard deviations obtained from the repeated measures described in
\S~2.

The completeness is larger than 80\% for $V\le 25.0$ for all the
considered subsamples. Since we limit the following analisys to $V\le
24.0$, we are well within a nearly-constant completeness regime, more
than one magnitude brigther than the limit at which the completeness
begins to fall rapidly to zero (around $V\sim 25.5$). The derived
completeness function slightly overestimates the actual completeness,
because of the severe selection criteria applied to the observed
sample, nevertheless it gives a clear indication of the magnitude range
in which the completeness is high and essentially constant with
magnitude. Analogously, the photometric uncertainty as estimated from
the artificial stars is in principle slightly larger than the actual values,
since in the observed sample each magnitude is the
average of at least three independent measures. However it provides an
independent check on the photometric accuracy and very stringent
constraints on the {\em distribution} of photometric errors, for
instance on the symmetry between the errors that make a given star
bluer or redder than it was in input (see below).

\subsection{Blue Main Sequence stars}

In Fig.~2 we compare the observed global CMD (stars from all the
cameras) with a CMD with the same number of stars randomly extracted
from the Similar Sample of artificial stars.

The synthetic CMD in the right panel of Fig.~2 was obtained as
follows: (a)  each extracted star was assigned the color of the
cluster ridge-line at the corresponding $V_{\rm input}$ by spline
interpolation; (b) the corresponding $(V-I)_{\rm output}-(V-I)_{\rm
input}$ (positive or negative) was added to the assigned color; (c)
$V_{\rm input}$ was substituted by $V_{\rm output}$.  In this way we
reproduce the effects of observations + data reduction on a parent
population of stars that were exactly placed on the cluster
ridge-line, as in \citet{n288}.  In both panels we also show for
reference, the ridge-line of the global sample shifted by $-0.05$ mag in
$V-I$.

The comparison between the observed and synthetic CMDs shows that the excess of 
stars to the blue and to the red of the MS described above cannot be due to 
observational effects but are intrinsic to the observed sources, instead. 
Hence, the observed population must include (1) a non-negligible fraction of 
real binary systems that appears as the usual parallel sequence to the red of 
the MS, and (2) a significant population of stars scattered blueward of the 
MS in the whole sample from $21.0\le V\le 24.0$. The origin of this 
anomalous blue MS population is unkown.

In order to quantify the frequency of the blue MS stars, we
study, in Fig.~3, the distribution of color deviation from the cluster
ridge-lines ($\Delta (V-I)$), separately for each subsample (camera),
in the range where the effect is more clearly visible ($22.0\le V\le
24.0$).  To disentangle the various components of the MS we subtract
from the observed distributions gaussian curves with $\sigma$ equal to
the typical error in color in that magnitude range. These gaussians
well fit the core of the distributions as shown by the small residuals
in the few central bins.  The residuals of the subtractions having
$-0.2\le \Delta (V-I) <0.0$ are divided by the total number of stars
in the distribution to obtain the fraction of blue stars ($B$) and the
same is done for $0.0<\Delta (V-I) \le 0.2$ residuals ($R$).  The
population of blue MS stars clearly emerges in all the
samples as a clear bump in the residuals containing $\sim 20$\% of the
whole MS stars in the considered magnitude range.

It is interesting also to compare, in the same magnitude range, the
symmetry properties of the observed color deviations from the ridge line
with those derived from the artificial stars shown in Fig.~2. In this
case we call $B$ the number of stars having negative $\Delta (V-I)$ and $R$
the number of those with positive $\Delta (V-I)$. The total observed
sample has ${{B-R}\over{B+R}}=0.12 \pm 0.02$, clearly indicating an
excess of blue stars, while the deviations of the artificial stars
sample are extremely symmetric, having ${{B-R}\over{B+R}}=0.01 \pm
0.01$. The difference between the two values is significant at the 4.4
$\sigma$ level. Analogously we find $B/R=1.26\pm 0.06$ for the observed
sample, to compare with $B/R=1.01\pm 0.02$ of the synthetic one.

\subsection{A possible role of differential reddening?}

The interstellar extinction toward NGC~2808 is quite large for a halo
globular cluster \cite[$A_V=0.68$ mag][]{h96}, hence we address the
question of wheater small scale variations of reddening could lead to
the observed difference in color among MS stars (blue vs. ``normal''
MS). There are two pieces of observational evidence
against this hypothesis: first, Blue MS stars are evenly spread over
the whole field and do not show any particular clustering property
(see also Fig. 4 below: however, this is not a strong argument, as the interstellar
dust is distributed in sheets and filaments, and may vary on angular
scales of arcseconds);  second, and more relevant, the color spread
along the MS occurs just for $V\ge 21.0$ (see Fig.~1 and Fig.~2).
If differential reddening were the cause of the color spread we would
observe it in any region of the CMD and, in particular, it should be
very evident in the nearly vertical portion of the sequence around the
Turn-Off point ($V\simeq 19.5$--20.0). On the contrary the MS is
clearly unimodal there and displays a color distribution fully
compatible with what expected from the photometric errors alone (see
Fig.~2).

Given the above evidence, we conclude that the anomalous population
of Blue Main Sequence stars is not due to the presence of differential
reddening in the (small) field considered here.

\subsection{Radial distribution}

While the sampled field is so small ($<2\farcm 6 \times 2\farcm 6$)
that the contamination by any possible system other than the cluster
itself should be negligible, the membership of the blue MS stars to
NGC~2808 must be established.  This is particularly relevant in the
present case, since the cluster may be physically associated with the
newly discovered Canis Major galaxy \citep{martin} and, in any case,
is projected in the same area of sky of this stellar system whose
optical CMD display a narrow and well defined MS, with a TO around
$V\simeq 19.0$ \citep{bell,david}.  The only way to assess this point
with our data is to compare the radial distribution of blue MS stars
with that of ordinary MS cluster stars.

If blue MS stars are cluster members they must be distributed as any
other star of NGC~2808; on the contrary, if they belong to a system
with a much larger size (as Canis Major or any known Galactic
component) their spatial distribution should be approximately uniform
over the observed field.

The results of this test are shown in Fig.~4. In the upper panel we
show the adopted selection, stars to the blue and to the red of the
shifted ridge-line, while the cumulative radial distributions are
displayed in the lower panel.  The distribution of blue MS stars is
indistinguishable from that of normal MS cluster members and is
strongly different from an uniform distribution over the considered
field. A Kolmogorov-Smirnov test yields that the probability that the
blue MS sample is drawn from a uniformly distributed population is
$P=3.4\times 10^{-10}$. Hence blue MS stars must be associated to
NGC~2808.

In conclusion, we have observationally established that $\sim 20$\% of
genuine MS stars of NGC~2808 are intrinsically bluer than the main
locus of most of the other MS stars of the cluster. In the following
sections we suggest an interpretation of this unexpected and puzzling
observational result.

\section{The theoretical models and the simulations} 
We computed new HB models, in addition to those presented in 
\cite{dantona-caloi2004}, namely the models with $Y=0.40$. The core mass at the 
helium flash for GC ages (12--13 Gyr) is in this case $M=0.465\,$\msun, 
compared to $M=0.495\,$\msun\ for $Y=0.24$.  Extreme HB stars in which only core 
helium burning is active, with the H-shell not contributing to energy generation 
are about 0.4 mag magnitude fainter for $Y=0.40$ than for $Y=0.24$. HB stars 
with $Y=0.40$ can extend to the faintest blue clump of the HB of NGC~2808, by 
assuming a H-envelope of the order of $10^{- 3}\,$\msun\ around the helium 
burning core of 0.465\,\msun (ZAHB luminosity $\sim 1.16\,$\Lsun, $T_{\rm eff} 
\sim 31,500\,$K, see Fig. 5). Synthetic models for the HB are computed following
the procedure described by \cite{dantona-caloi2004}.

For the MS study, we computed stellar models down to 0.3\Msun,  
to construct isochrones for 
metallicity $Z=10^{-3}$ and helium content $Y=0.24,~0.28,~0.32$ and 0.40, and 
for metallicity $Z=2\times 10^{-3}$, for $Y=0.24,~0.30$ and 0.40.  The input 
physics is described in D'Antona and Caloi (2004). Models for $M<0.4\,$\msun\ 
are built by making use of a new equation of state based on \cite{saumon, 
rogers2001, stolzmann1996, stolzmann2000} computations.

Synthetic models for the cluster main sequence, turnoff and subgiant population 
are built by interpolating among the isochrones with various assumptions about 
the helium content. We fix the age of the cluster, and the stellar mass is 
randomly extracted from a power law mass function (results will be 
shown for an exponent --1.5, in the notation in which Salpeter's 
index is --2.3; in any case, the choice of the mass distribution 
has no influence for the specific problem of this work). The 
number of stars for each chosen helium content is fixed a priori, and the 
resulting color distribution is compared with the observations, to choose the 
most appropriate ones. After each determination of magnitude and color, we 
apply an error extracted from a normal distribution with standard deviation 
chosen according to the observations.

In running the simulations of the MS color distribution we employed the 
following procedure: we first took the distribution of a sample of stars with 
standard helium abundance $Y=0.24$, then we added stars with increasing helium 
content, in order to reproduce reasonably the observed distribution in color. 
Our first aim is to reproduce the MS, including the blue MS stars. At the same 
time, we have also to worry about the observed HB morphology. Therefore, we 
will finally adopt helium distributions $N(Y)$ such that they are consistent 
both with the MS colors and with the two main features of the HB, the gap at 
the RR Lyrae and the distribution of stars in the EBT1 and EBT2+EBT3 blue 
clumps (we follow the definitions by \cite{bedin2000}). Of course, any $N(Y)$ 
which reproduces the HB but not the MS colors is not allowed, and viceversa. We 
ran simulations for both $Z=10^{-3}$\ and $2 \times 10^{-3}$, finding similar 
results for the MS color distribution.  

We show in Fig. \ref{f5} a simulation of the turnoff and MS, made according to 
the above described method, in which the N(Y) follows the distribution given in 
Fig. 9, which will be discussed in Sect. \ref{3.3}.  We superimpose the 
isochrones of 13Gyr, $Z=2 \times 10^{-3}$. A look at Fig. \ref{f5} shows an 
important feature. The theoretical isochrones with variable helium and fixed 
age converge at the TO region. This implies that, if helium variations are 
responsible for the blue main sequence subsample, significantly bluer than the 
MS ridge line and fainter than the TO, the intrinsic dispersion of the data at 
the TO should be negligible and due just to photometric errors. Examination of 
Fig.~1 and Fig.~2 shows that this is indeed the case.  The TO region in the CMD of 
NGC~2808 ($19.5<$V$<20.5$), is vertical and extremely narrow in color, and 
essentially no blue outlier is detectable.

\subsection{The distribution in color: the MS}
\label{3.3}
In Fig. \ref{f6} we show the color distribution corresponding to the simulation 
in Fig. 6, in the magnitude range $5.5<M_v<7.5$. In order to reproduce the MS 
observations, it would suffice to consider simply two helium values: $Y$=0.24 
for the vast majority of stars ($\sim 80$--85\%), and $Y \sim 0.40$ for the 
remaining fraction. However, in the simulation, we do not assume a unique 
Y=0.24 for this main fraction of stars, because of the necessity of reproducing 
the HB morphology, but the $N(Y)$\ shown in Fig. 9 for $0.24 \leq Y \leq 0.29$. 
The small additional scatter in the MS is consistent with the observed color 
distribution having observed standard deviation $\sigma$=0.03, which means that 
it remains hidden within the photometric errors. This follows from the fact 
that, in the models,

\begin{equation}
\Delta(V-I)/\Delta Y \simeq -0.438
\label{eq1}
\end{equation}
Thus, the color of stars having, e.g., $Y$=0.275, is shifted, with respect to a 
MS with $Y$=0.24, only by --0.015mag. 

According to Eq. \ref{eq1}, the blue MS can be reproduced only by adding a 
group of stars (from $\sim15$\% to $\sim22$\%, see Fig. \ref{f3}) with 
$Y=0.40$. This additional group of stars is absolutely necessary: in fact, had 
we adopted a more or less uniform spread in helium, it would have produced too 
many stars at $\delta(V-I) \simeq - 0.05$\ and not enough stars at $\delta(V-I) 
\simeq -0.10$.  The specific value $Y=0.40$\ should not be taken as a precise 
quantitative estimate, as our models suffer from at least one important 
uncertainty: the color versus \teff\ relationships are based on models by 
\cite{bessell-castelli-plez1998}, which are built with normal helium content.

\subsection{The distribution in color: the HB}
\label{hb}
The $N(Y)$\ distribution illustrated above and displayed in Fig.~9 is chosen in 
order to reproduce the HB morphology, at a first-order approximation. In 
particular it allows to obtain the observed lack of HB stars in the RR Lyrae 
region and the distribution of stars of the clump EBT1 \cite{bedin2000}. In 
summary: (1) $\sim 15$\% of the stars has $Y=0.40$; these stars account for the 
blue MS population and populate the EBT2 and EBT3 clumps of the HB; (2) $\sim 
50$\% of the stars have $Y=0.24$, and (3)$\sim 35$\% of the stars have $Y$\ in 
the range $0.26< Y <0.29$. The addition of the latter stars ---not 
required by the observed MS morphology--- does not significantly worsen the MS 
fit, as discussed above.  Fig. \ref{f7} shows that it also reproduces quite 
satisfactorly the HB general morphology.  Comparison of the full HB simulation 
is made with the \cite{bedin2000} observations.

The interesting result from the global simulation is that the reproduction of 
the MS color distribution, coupled with a good description of the whole HB 
morphology, naturally leads us to the conclusion that about 15--20\% of the 
cluster stars has a helium content $\simeq 0.4$, and shows up {\it both} in the 
MS color distribution (Figure \ref{f6}, the bluest MS stars), {\it and} in the 
HB, where it shows up in the two blue HB tails, EBT2 and EBT3.

Notice that we can not reproduce the gap between EBT2 and EBT3, but only the 
total number of stars. However, it is not straightforward to obtain the correct 
star distribution in these two HB regions. If we interpret observations in 
terms of mass distribution of stars with $Y = 0.40$, we find that the the 
population in EBT3 arises from only one mass, of about 0.466 \msun, which 
evolves vertically (see Fig. 5) more or less at constant pace, covering the 
observed faintest blue region. A spread in mass necessarily fills up the gap 
between EBT3 and EBT2\footnote{\cite{lee-demarque2005} attribute the EBT2 and 
EBT3 clumps to two populations differing in helium content, but they do not 
explain why each of these should have practically a unique evolving mass.}. For 
the latter group of stars the situation is not very different: the bulk of it 
is confined between 0.48 \msun\ and 0.49 \msun. A possible interpretation is to 
consider EBT2 stars as deriving from red giants which develop the helium flash 
at the tip of the giant branch, that is, ``normal'' HB stars. With an evolving 
giant of about 0.63 \msun, the mass loss involved is of about 0.15 \msun\ and 
the envelope mass of about 0.015 \msun, sufficient for a standard helium flash. 
Larger mass losses would give origin to early or late hot flashers, as defined 
and discussed by Castellani \& Castellani (1993), D'Cruz et al. (1996), Brown 
et al. (2001), that is, stars in which the core-helium flash takes place, 
either at the beginning or later, along the white dwarf cooling sequence. In 
this respect, a fundamental piece of information comes from the observations by 
Brown et al. (2001) and Moehler et al. (2004). In particular, Moehler et al. 
(2004) obtained spectra of EBT3 members, finding evidence both of the high 
surface temperatures (35000 K and more) and of the large surface abundance of 
helium and carbon, expected as a consequence of the mixing during the late 
core-helium flash (Sweigart 1997). This appears to decisively settle the 
question as to the origin of these stars.

On the other hand, the simulation shown in Fig. \ref{f7} has a problem: we have 
to assume a mass loss of $\sim 0.195\,$\msun\ during the RGB evolution, in 
order to reproduce the red side of the HB with $Y=0.24$ models and the clump 
EBT1, with the assumed $N(Y)$. Then the mass loss has to be reduced to only 
$\sim0.15\,$\msun\ to account for the extreme blue tails with $Y=0.40$. If the 
mass loss regime does not change too much with helium content, the $Y=0.40$ 
evolving giants should leave the giant branch before reaching the conditions 
for the core-helium flash. On the contrary, the difference between the fraction 
of blue MS stars and the fraction of EBT2+EBT3 stars may amount to $\sim 5$\%, 
according to our and \cite{bedin2000} observational estimates. This is just one 
of the many questions and problems that this exploratory investigation leaves 
open.  If the helium content of the blue MS could be reduced to $Y\sim0.35$, 
the mass loss problem would be practically solved. We have already noticed the 
uncertainties in the color-\teff\ transformations, which might depend on $Y$, 
not taken into account in the model atmospheres.

At this stage we can not pretend to explore all the details of the HB 
morphology of this cluster, as there is a number of parameters (mass loss 
included) which certainly influence it (Rood 1973, Fusi Pecci et al. 1993, Fusi 
Pecci and Bellazzini 1997).  Nevertheless, we stress that the quantitative 
reproduction of the main anomalies, both of the MS (blue MS stars) and of the 
HB (red HB, gap at the RR Lyrae, blue HB with extreme blue tails) indicates 
that the proposed interpretation in terms of helium variations is one of the 
most promising proposed till today.

\section{The star formation events in NGC~2808: a new global scenario}

The result of this work provides a new piece of information on the early stages 
of formation of stars in GCs. In \cite{dantona-caloi2004} we had hypothesized a 
more or less continuous star formation starting after the period of supernovae 
II explosions. After a few $10^7$ yr the slow winds from massive AGB stars 
would begin to collect helium enriched material in the cluster core and form stars.

Fig. \ref{f8} shows the comparison between the $N(Y)$ suggested in 
\cite{dantona-caloi2004} and that derived in this work. The latter seems to 
depict a situation very different from the continuous star formation.  However, 
it should be preferred since the MS color distribution is a much better 
determinant than the HB distribution for deriving $N(Y)$. The HB distribution 
depends on many parameters other than $Y$ so it can never yield a unique 
$N(Y)$.

The HB distribution does suggest that NGC~2808 probably underwent three main 
stages of star formation.  The {\it first} one is easily identified with the 
main burst of star formation, and is now probed by the low mass stars with 
primordial helium content ($Y=0.24$), which constitute $\sim 50$\% of the MS 
and populate the red HB clump.  The {\it second} burst in chronological order 
results in the 15--20\% of population at $Y=0.40$.  Finally, there is a {\it 
third} stage of star formation producing $\sim 30-35$\% of the stars with 
helium larger than the primordial value up to about $Y=0.29$. These stars 
populate the HB clump EBT1.

The reason why the stars from the second burst of formation have the highest 
$Y$\ is simply that the helium content in the AGB envelopes increases with the 
initial stellar mass, due to the higher efficiency of the second dredge up 
\citep{ventura2002}. \cite{ventura-dantona2005b} estimate $Y=0.32$ for their 
most massive models (6.5 \msun) for metallicity $Z=10^{-3}$. This value can be 
considered conservative, as their models do not include any kind of 
overshooting, which may easily increase the helium content to the required 
value of $Y=0.40$. \cite{lattanzio2004} find $Y\sim$0.36 from the evolution of 
a 6\msun\ with $Z=0.004$. In addition the episodes of third dredge up may help 
to increase the helium content in some models \citep{ventura-dantona2005}. 
Notice also that $Y\sim 0.35$ would reduce the problems in the reproduction of 
the blue tails EBT2 and EBT3 (see Sect. \ref{hb}). Therefore we suggest to 
identify the population with extreme helium abundance ($Y\sim0.40$) as the 
stars born from the winds of the {\it most massive} AGB stars, which evolve 
$\sim 50$ Myr after the first burst of star formation.

The third group of stars may be the result of star formation from the winds of 
somewhat smaller mass AGBs.  If we rely on the AGB models by 
\cite{ventura2002}, the best mass range for these objects is from $\simlt 
4.5\,$\msun\ to $\simgt 3.5$, which evolve at ages $\sim 100$--150\,Myr after 
the first burst of star formation. Hence, the ``intermediate'' helium rich 
stars are the third stage of star formation in chronological order. 

We are aware that this scenario requires an initial mass function of the first 
burst of star formation strongly weighted towards intermediate mass stars, in 
order to get the necessary amount of helium enriched ejected matter.  This 
aspect of the problem has been extensively discussed in \cite{dantona-caloi2004}.

\subsection{The possible role of delayed Type II Supernovae explosions in binaries}

In our scenario there are no stars formed with helium $0.29 < Y < 0.40$. 
Naively one would expect that, as the cluster ages, the mass of the AGB stars 
decreases as does the helium abundance in the winds and in the newly born 
stars. If the formation of new stars is continuous why do we not observe stars 
with all $Y$ values between 0.29 and 0.40? We briefly consider three possible 
answers.

First of all our modeling of AGB stars is not so secure to exclude 
discontinuities in the helium content of the AGB ejecta. In fact, some of the 
models \citep[][Fig. 4]{ventura2002} show a non-linear correlation between the 
evolving mass and the helium yield. Depending on the efficiency of star 
formation, on the modalities of mixing of the intracluster material ejected 
from the AGBs, and on the initial mass function of the AGB stars, there could 
be a relative rarity of stars in some helium abundance ranges.

Secondly, it is possible that external events are needed to trigger the star 
formation phase. In this case, after a first trigger when $Y=0.40$, the gas 
from the AGB ejecta goes on accumulating in the cluster central regions for a 
long time (several tens of million years) until the second trigger occurs when 
the average helium content in the gas is reduced to $Y\sim 0.28$.

A third possibility requires a bit more of explanation. The uniform metallicity 
of GC stars is generally attributed to the fact that the explosion of type II 
supernovae do not pollute the gas from which the stars are formed, probably 
because they bring an end to the first burst of star formation by clearing from 
the gas the intracluster medium.  Type II supernova continue until all single 
stars with $M> M_{\rm up}$ (where $M_{\rm up}$\ is the maximum mass for white 
dwarf formation) have exploded. At that time the low velocity winds from the 
evolving AGB stars begin to accumulate in the core, and give origin to the 
second stellar generation with $Y\sim 0.40$. This can last until {\it there are 
new type II supernova explosions from evolving binaries} which again expel the 
gas from the cluster.

Mass exchange between the components of primordial binaries can push the mass 
accreting component with initial mass $M < M_{\rm up}$, to a mass $M > M_{\rm 
up}$ converting it to a Type II supernova progenitor. The timing of such 
supernovae is set by the time it takes the primary with mass $M_1$ to reach the 
AGB. They can continue until mass exchange is no longer capable of pushing the 
accreting star above $M_{\rm up}$. To place a limit on this time scale, we 
assume that all the primary mass---apart from its white dwarf remnant---is 
transferred to the secondary, and that the secondary initially has a mass 
comparable to the primary. We obtain:
\begin{equation} M_1 > (M_{\rm up} + M_{\rm WD})/2 
\label{mass}
\end{equation}
 
\noindent For $M_{\rm up}=7\,$\msun\ and $M_{\rm WD}=1$\msun, the primary mass 
must be larger than 4\,\msun. While a more precise estimate of this value is 
needed, but it is not inconsistent with our claim that we expect the winds to 
begin again a third star formation epoch when $M_{\rm AGB} \sim 4.5\,$\msun.

The hypothesis of late supernovae II explosions may be an important 
ingredient in the GC evolution. In particular, $M_{\rm up}$ decreases for 
smaller metallicities: this may explain the smaller degree of chemical 
abundance varations in the lowest metallicity clusters, as the late Type II 
supernova from binary evolution will last in the cluster for a longer time (see 
equation \ref{mass}), preventing star formation for such a long time that the 
conditions for producing a new stellar generation, in addition to the first 
burst, become more difficult.

\section{Discussion and conclusions}

We have presented HST observations of the MS of the GC NGC~2808, which provide 
important new evidence concerning the early evolution of GCs.  15--20\% of the 
MS stars is bluer by $\delta(V-I) \sim -0.1$\ than the bulk of the MS stars 
(80--85\%) whose colors follow a normal distribution with standard deviation 
consistent with the observational errors. The blue MS are most simply 
interpreted as a group of stars substantially coeval with the other cluster 
stars ($\Delta t \leq 50\,$Myr), but having helium abundance $Y \simeq 0.40$. 
The two most obvious anomalies of the CM diagram of NGC~2808, namely {\it i)} 
the blue MS stars and {\it ii)} the very peculiar morphology of the HB, can be 
simultaneously reproduced by means of the variations of the single parameter 
$Y$.

This is not the first time that the helium content is proposed to play a role 
in the MS morphology. In \ocen, the presence of a high helium MS has been 
suggested \citep{bedin2004,norris} as one of the possible components needed to 
explain the complexity of this cluster, where multiple turnoffs and red giant 
branches have been detected \cite[see,][and references 
therein]{leeomega,pancino, ferraro2004,bedin2004,sol05}.

Recently, on the basis of the magnitude difference between the turnoff and the 
red giant bump, \cite{caloi2005} have proposed a difference in the global 
helium content of M3 and M13, and have shown that such interpretation provides 
subtle and consistent differences in the turnoff morphologies. The higher 
helium in M13 can also account for its much bluer HB.

In NGC~2808, {\it the evidence for a high helium population is much more 
compelling than in the other quoted cases}, as the cluster could have been 
considered a standard ``simple stellar population,'' having no metallicity 
spread \cite[in terms of Fe and alpha elements, see,][and references therein]{cbc04}
\footnote{In addition to the 20 stars examined in \cite{cbc04},
Carretta, Bragaglia and Gratton (private communication) have now 
reduced 123 high resolution spectra of red giants in NGC 2808: according to this new
analysis, the maximum metallicity spread allowed by the data amounts 
to a few hundreths of dex ($\pm 0.03$).}, contrary to the \ocen\ case. 
In addition, the evidence is based on relative photometric measurements of the 
MS, and not on comparison of different clusters CM diagrams, as in the case of 
M3 and M13.

The case for cluster self-enrichment from the ejecta of massive AGB stars, is 
strengthened by the present discovery of a high helium population \cite[see 
also][and references therein, for other possible signatures of early self-
enrichment]{carretta2003,cbc04}. The enrichment in helium is one of the more 
robust predictions in the computation of the stellar yields from AGB stars and 
is much less model dependent than the processing of the CNO and of the other 
proton-capture elements \citep{ventura2001}. The precise abundance of the high 
helium population depends on the color--\teff\ transformations, and we urge the 
computation of model atmospheres with the cluster chemical abundances and 
taking into account helium enhancement.  

While it is common knowledge that helium abundance affects many phases of GC 
evolution, it has generally been ignored in the studies of globular clusters in 
the last decades. Our results suggest that helium abundance may play a major 
r\^ole in determing differences in the populations of different clusters {\em 
and} within the population of a given cluster, hence its possible influence 
cannot be neglected.

We propose a possible scenario that simultaneously accounts for the anomalus MS 
and HB observed in the CMD of NGC~2808 in terms of variations of helium 
abundance. Our scenario suggests {\it three main stages of star formation} in 
NGC~2808, all of them globally lasting not more than 200\,Myr: a {\it first} 
burst at the Big Bang abundance $Y=0.24$, a {\it second} one from the gas of 
the most massive AGBs at $Y\simeq 0.40$, and a {\it third} one from less 
massive AGBs, having $Y$\ from $Y\sim 0.29$ down to $Y\sim 0.26$ and a $Y$\ 
distribution peaked at $Y\simeq 0.27$. In this framework, the late binary 
supernova II explosions may play a major r\^ole in producing the postulated 
discontinuity in helium abundance between $Y\sim 0.40$ and $Y\sim 0.29$.

We must be careful not to over interpret our results. The existence of the high 
helium ($Y\sim 0.40$) population seems robust since it is based both on the MS 
spread and HB morphology. The intermediate helium population is inferred only 
from the HB morphology. Some support is given by the fact that the reddest 
stars of the blue HB are slightly more luminous than the red clump stars. The 
different luminosities result naturally from the slight difference in the $Y$\ 
of the red clump and the red side of the blue HB \citep{dantona-caloi2004}. 
However, as also our discussion shows, the interpretation of HB morphology is 
always intertwined with the question of mass loss. Hence, independent 
observational evidence is required to check the actual existence of this 
intermediate-$Y$ population. In this respect, we consider the cluster RR Lyr 
variables. Recently their number has been raised from 2 to 18 \citep{corwin}. 
The mean periods of RR$_{ab}$ and RR$_c$ pulsators are $\sim$0.563 d and 0.30 
d, respectively; the ratio $N_c/N_{ab}$\ is about 0.44. The latter value would 
favor a classification of RR Lyrs as Oosterhoff II type, while the mean periods 
indicate an Oosterhoff I type. The mean periods suggest for these variables a 
luminosity close to that in other OoI GCs like M3 and M5. Since for the stellar 
population in the latter clusters (at least for its bulk) there is no 
indication of a larger than cosmological helium abundance, the RR Lyr periods 
observed in NGC 2808 may be considered an indirect hint for a close to normal 
value of $Y$. This is not a surprise, as the RR Lyr could be the bluest 
extension of the red clump population, which has a normal $Y$, but 
unfortunately, does not provide us any further support for the existence of the 
intermediate $Y$ population. 

Our observations suggest a possible observational strategy to search for the 
effects of early nucleosynthesis in GCs. Why a clear main sequence 
bimodality has been
discovered so far only in this cluster and in \ocen? First of all, the signal of 
an extreme helium rich population is very weak, and requires deep and accurate 
photometry in uncrowded regions of massive globulars.  
Only by means of these observations we may reveal the same signature observed 
in NGC~2808, e.g., an anomalous blue population in the sloped part of the MS 
with  no counterpart in the TO region. Obvious candidates may be clusters with 
multi-modal HB morphology as NGC~1851, NGC~6441 and NGC~6388, but the first
cluster has no published HST photometry, and the test may 
result unfeasible for the latter two clusters, because they are significantly 
affected by differential reddening that may smear out a subtle signal as the 
one put in evidence in the present analysis \citep{sosin}. In any case, the 
spatial resolution and photometric accuracy required to identify a minor 
component of anomalous MS stars within a globular cluster are quite challenging 
and it may turn out that the interesting clusters accessible with this 
technique are very few. The individual history of each cluster will also affect
the consistency of the extreme population: if it is as low as, say, 10\% of the
total, it can easily pass unnoticed. Notice that even HST main sequence 
photometry is useless to discriminate stars with helium contents from $Y$=0.24 to
$Y$=0.28, although such a small spread corresponds to large CNO and
sodium anomalies \citep[see, e.g. the models by][]{ventura2002}s.

Finally, we want to draw the attention of the readers to a few interesting 
facts that may be intimately related with the above results and discussion. 
First, it is worth noting that the extreme blue HB becomes fainter and fainter 
with increasing helium abundance and the ZAHB loci at various $Y$ intersect 
each other (see Fig. 5).  As a consequence the group of HB stars fainter than 
the ``normal'' ZAHB detected, e.g., in \ocen\ and in NGC~2808 with UV 
observations \citep{dcruz00,brown-sweigart2001} could perhaps be intepreted as 
belonging to a sub-population having an helium abundance higher than those 
populating the ``normal'' ZAHB, in agreement with the suggestions put forward 
to explain the other main branches \citep{norris}. 

Second, \citet{fpetal93} pointed out a correlation between the extension of 
blue HB tails and the integrated absolute magnitude of globular clusters, 
particularly evident when the sample is restricted to clusters with 
intermediate metallicity ($-1.9< \feh <-1.2$) where, as discussed in that 
paper, the sensitivity to any second parameter effect is better visibile. If 
self-enrichment (and multiple bursts of star formation) can increase the helium 
abundance of subsamples of stars, it seems quite natural to imagine that more 
massive (and/or more concentrated) clusters would be more effective in keeping 
the helium-enriched ejecta of AGB stars, and, in turn, more efficient in 
producing stars populating the bluest part of the HB.

Third, \citet{bur04} found that there is a marked difference in nitrogen 
abundance for the stars in the old M31 GCs relative to those in the oldest 
Galactic GCs. As they noted, the M31 GCs for which very strong NH absorption 
has been detected have absolute luminosities ranging from $M_V=-11$ to $-9$, 
i.e., are very bright and massive. Therefore, it can be conceived that this 
peculiar enhancement is indicative of a corresponding higher helium abundance 
as, within the framework described here, nitrogen and helium enhancements could 
result from the self-pollution occuring during the first stages of cluster 
life.

In conclusion, while we are {\em not} claiming that the whole variety of not-
fully-understood phaenomena encountered in Galactic globular clusters can be 
explained in terms of variations in helium abundance, the case of NGC~2808 
strongly indicates that such variations do occur in real clusters and early AGB 
nucleosynthesis may play a significant r\^ole in the evolution of, at least, 
some of them. 

\clearpage

\clearpage

\begin{figure}
\centering
   \resizebox{7.8cm}{!}{\rotatebox{0}{\includegraphics{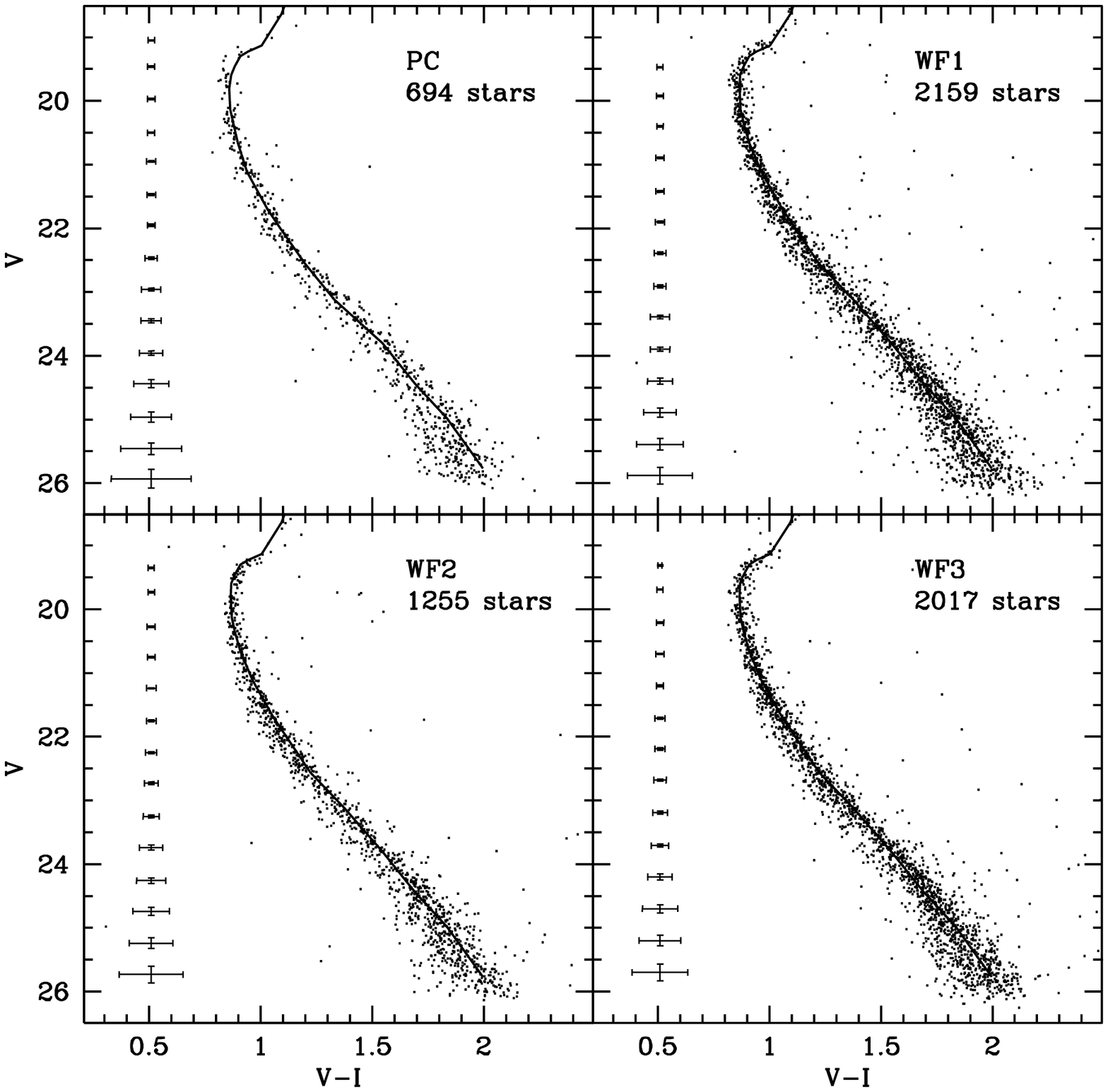}}}
\figcaption[fig1.ps]{CM diagrams from the four WFPC2 cameras. The error bars
have been obtained from the artificial star experiments 
($\sigma_V =V_{\rm output}-V_{\rm input}$ and $\sigma_{V-I}
=(V-I)_{\rm output}-(V-I)_{\rm input}$). The ridge lines of each sample are
over plotted.
\label{f1}} 
\end{figure}

\begin{figure}
\centering
   \resizebox{7.8cm}{!}{\rotatebox{0}{\includegraphics{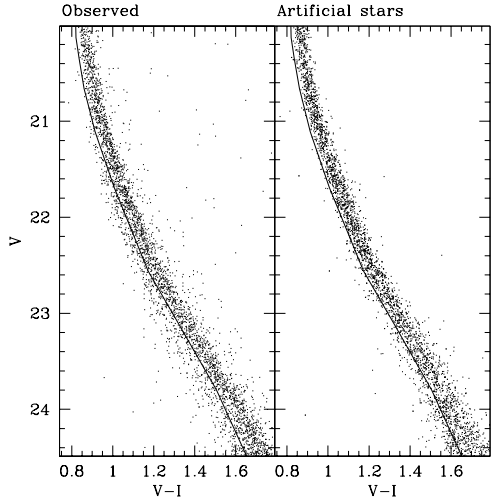}}}
\figcaption[fig2.ps]{Left panel: zoomed view of the total CMD (4 chips).
Right panel: synthetic CMD obtained from the artificial star experiments. 
The MS ridge line over plotted to both diagrams has been shifted 
by -0.05 mag in color.
\label{f2}} 
\end{figure}

\begin{figure}
\centering
   \resizebox{7.8cm}{!}{\rotatebox{0}{\includegraphics{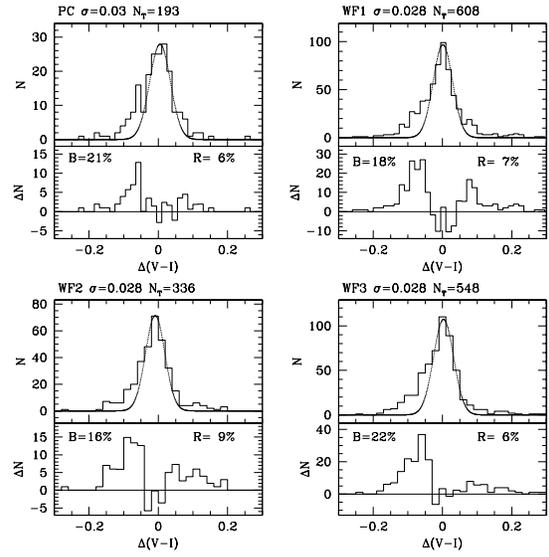}}}
   \figcaption[fig3.ps]{The distributions of the color deviations from
   the ridge line for the stars having $22.0<V<24.0$ in the four
   samples are shown as histograms in the upper part of each
   panel. The dotted lines are gaussian curves fitting the core of the
   observed distributions. The adopted $\sigma$'s are the typical
   $\sigma_{V-I}$ obtained from the artificial star experiments. The
   lower part of each panel shows the residuals of the subtraction of
   the gaussian model from the observed distribution of deviations. The
   fractions of residual stars having $-0.2\le \Delta(V-I)<0.0$ (B)
   and $0.0< \Delta(V-I)\le 0.2$ (R) are also reported.
\label{f3}} 
\end{figure}

\begin{figure}
\centering
   \resizebox{7.8cm}{!}{\rotatebox{0}{\includegraphics{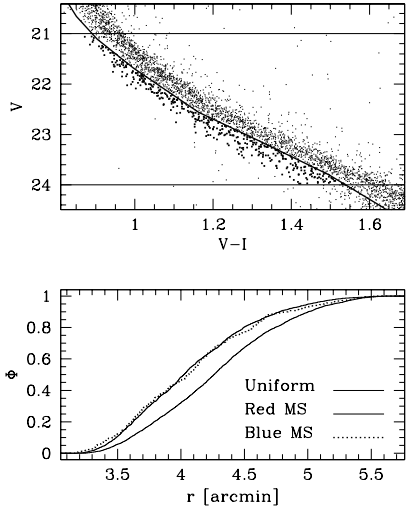}}}
\figcaption[fig4.ps]{Upper panel: selection of Blue MS (to the blue of the
shifted ridge line, large dots) and Red MS stars (to the red of the shifted 
ridge line). The horizontal lines enclose the magnitude range of the adopted
selections. Lower panel: cumulative radial distributions of Blue (dotted line) 
and Red (continuous line) MS stars.
\label{f4}} 
\end{figure}

\begin{figure}
\centering
   \resizebox{7.8cm}{!}{\rotatebox{0}{\includegraphics{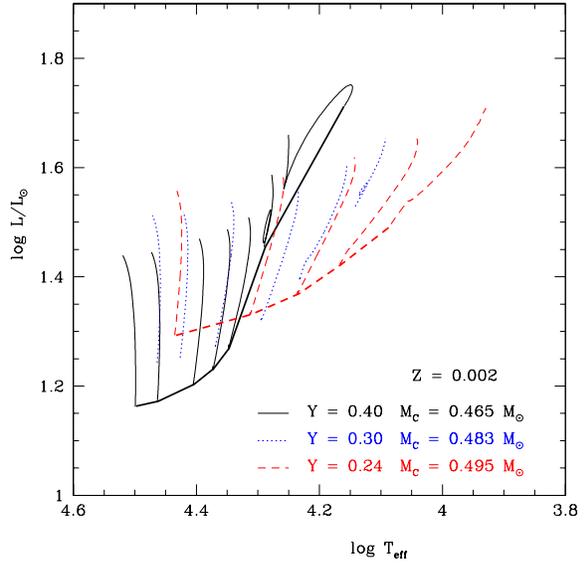}}}
   \figcaption[fig5.ps]{Evolutionary tracks for the hottest region of
   the HB. Tracks are terminated when the core helium content is
   reduced to $Y=0.10$, From left to right masses are: 0.5, 0.52,
   0.54, 0.56 and 0.58 \msun\ ($Y=0.24$); 0.486, 0.49, 0.50, 0.52,
   0.54, 0.56 \msun ($Y=0.30$); 0.466, 0.469, 0.48, 0.49, 0.50, 0.52,
   0.54\msun\ ($Y=0.40$). For $T_{\rm eff} \simgt 20000\,$K, the
   $Y=0.40$ ZAHB is less luminous than the ZAHB for $Y=0.24$, because of
   the reduced helium core mass.  At $T_{\rm eff} \simlt 20000\,$K, the
   contribution of the H-shell burning to the luminosity begins to be
   relevant, and the effect is more important for larger $Y$ (see, e.g.,
   the luminosity of $M=0.52$ and 0.54\,\msun\ for $Y=0.40$). Heavy continuous line:
   ZAHB with Y=0.40; heavy dashed line: ZAHB with Y=0.24.
   \label{hbfigure}} 
\end{figure}

\begin{figure}
\centering
   \resizebox{6.8cm}{!}{\rotatebox{0}{\includegraphics{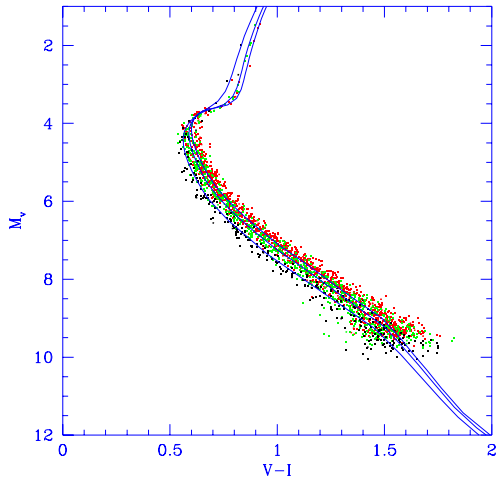}}}
\figcaption[ngc2808_1.eps]{
The three isochrones have age 13Gyr, $Z=2 \times 10^{-3}$, and $Y=0.24$, 0.30 and 0.40. 
The simulation based on these three isochrones has the $N(Y)$\ shown in Fig. 9.
The standard deviation in color $\sigma$\ added to the simulation
is fixed on the basis of the observational errors, which increase with the
magnitude.
\label{f5}} \end{figure}

\begin{figure}
\resizebox{6.8cm}{!}{\rotatebox{0}{\includegraphics{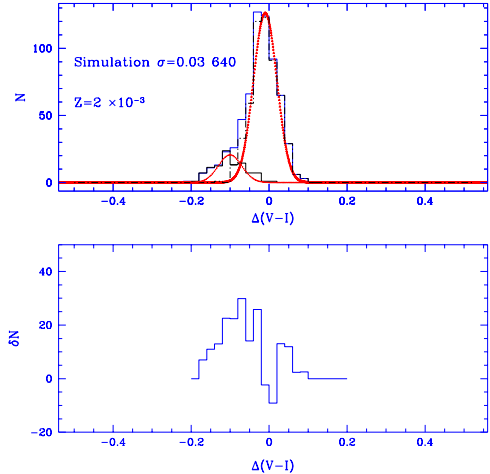}}}
\figcaption[ngc2808_3.eps]{The color distribution of the main sequence
stars in NGC~2808 in the field WF3. The stars are limited to the range
between
$M_v=5.5$\ and 7.5. The gaussian fit with the same standard deviation as the
data ($\sigma = 0.03$) shows an excess of stars at blue colors.  The
observed distribution, extended to the blue up to $-0.2$ mag from the
median line, requires a distinct population with high helium to be
reproduced. This population at $Y=0.40$, including about 15\% of the
total stellar population in this main sequence range, is shown
separately by the histogram at the left, fitted with a gaussian. The
dash-dotted histogram includes the stars with helium content up to
$Y=0.29$. The bottom
panel shows the differences between the simulation counts and the
gaussian expected values.
\label{f6}} 
\end{figure}

\begin{figure}
\centering
\resizebox{6.8cm}{!}{\rotatebox{0}{\includegraphics{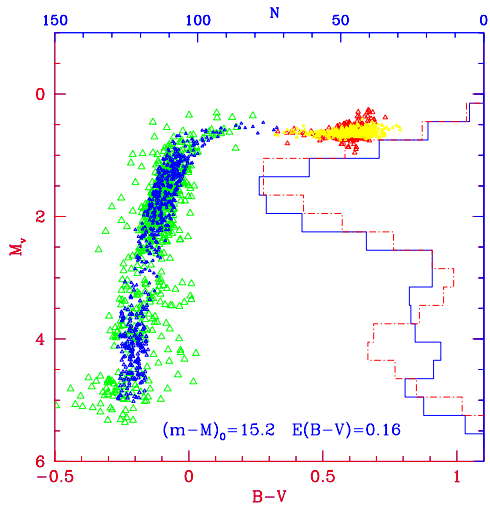}}}
\figcaption[ngc2808_6.eps]{ By assuming the same helium distribution
(Figure \ref{f8}) which reproduces the MS colors, and some additional
assumptions relative to the mass loss, we obtain a synthetic HB
distribution, compared here to the observed data for the cluster
NGC~2808 by \cite{bedin2000}.  The red clump is made up by stars with
$Y=0.24$, the blue clump EBT1 (the most conspicuous and luminous)
contains stars with $0.26<Y<0.29$, with a strong prevalence of stars
at $Y=0.27-0.28$; the clumps EBT2 and EBT3 contain the stars with $Y=0.40$.
 \label{f7}} \end{figure}

\begin{figure}
\centering
\resizebox{6.8cm}{!}{\rotatebox{0}{\includegraphics{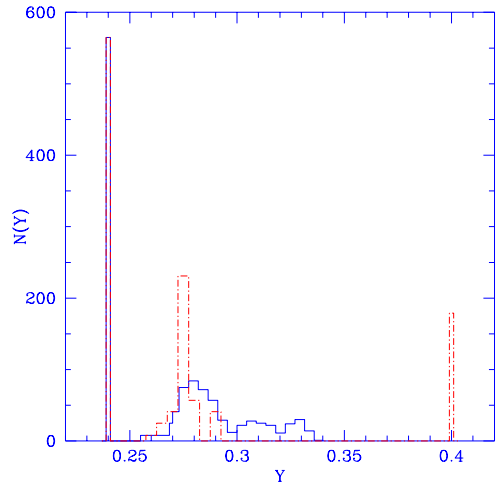}}}
\figcaption[ngc2808_5.eps]{ The helium distribution adopted for the
simulation shown in Fig. \ref{f6} and \ref{f7} is shown dash-dotted,
and is compared with the helium distribution predicted by D'Antona and
Caloi (2004) for the HB (full line histogram).
 \label{f8}} \end{figure}

\end{document}